\documentclass[aps,prl,reprint,twocolumn,showpacs,superscriptaddress,amsmath,amssymb,floatfix]{revtex4-1}
\usepackage{graphicx}
\usepackage{svg}
\usepackage{psfrag}
\usepackage{epsfig}
\usepackage[normalem]{ulem}

\begin{document}


\title{Control of superselectivity by crowding in three-dimensional hosts}


\author{Andrew T. R. Christy}
\affiliation{Department of Chemistry, Durham University, South Road, Durham, DH1 3LE, United Kingdom}
\author{Halim Kusumaatmaja}
\email[]{halim.kusumaatmaja@durham.ac.uk}
\affiliation{Department of Physics, Durham University, South Road, Durham, DH1 3LE, United Kingdom}
\author{Mark A. Miller}
\email[]{m.a.miller@durham.ac.uk}
\affiliation{Department of Chemistry, Durham University, South Road, Durham, DH1 3LE, United Kingdom}



\date{\today}

\begin{abstract}
Motivated by the fine compositional control observed in membraneless droplet organelles
in cells, we investigate how a sharp binding--unbinding transition can occur between
multivalent client molecules and receptors embedded in a porous three-dimensional structure.  
In contrast to similar superselective binding previously observed at surfaces, we have identified that a key effect in a three-dimensional environment is that the presence of inert crowding agents can significantly enhance or even introduce superselectivity. In essence, molecular crowding initially suppresses binding via an entropic penalty, but the clients can then more easily form many bonds simultaneously. We demonstrate the robustness of the superselective behavior with respect to client valency, linker length and binding interactions in Monte Carlo simulations of an archetypal lattice polymer model.
\end{abstract}


\maketitle

Superselectivity is the sensitive response of binding to the density of binding
receptors.  If a system exhibits superselectivity then there is a characteristic receptor
density at which the binding species changes
sharply from being mostly free to mostly bound \cite{martinez2011designing}.  This
effect is desirable whenever a clean switch is required 
in response to a well-defined trigger.  Applications include tumor treatment (where target
cells can be recognized by the anomalous expression of certain receptors on their
surface) \cite{carlson2007selective}, targeted drug delivery \cite{ZHANG20139728},
and self assembly in materials chemistry \cite{Huskens2004}.  Superselectivity is
also exploited by nature in cell signalling and immunology \cite{multivalencyuses} because
the step-like transition allows binding to be finely
controlled \cite{Dubacheva2015}.  Superselectivity relies on the binding species
being multivalent; it is the ability of multivalent particles to form many combinations
of multiple, weak connections that makes the binding sensitive to receptor concentration
while still providing a strong overall
interaction \cite{martinez2011designing,mammen1998polyvalent}. 

The archetypal model of superselective binding is a surface decorated with binding receptors
and a nanoparticle coated with ligands.  Binding is triggered when the surface density of
receptors is high enough for the nanoparticle to interact with many receptors
simultaneously \cite{martinez2011designing}.
A similar effect at a surface can be achieved when the binding species is a multivalent
polymer, where multiple binding sites along the polymer play the role of the
ligands grafted onto the nanoparticle \cite{Dubacheva2015}.

In contrast to existing work on superselective binding at surfaces, here we examine
whether superselectivity can also be observed for receptors embedded in a porous three-dimensional
(3D) structure.  Significant motivation for this work comes from membraneless droplet
organelles.  These cellular substructures are formed by mechanisms
resembling liquid--liquid phase separation (LLPS) \cite{jain2016atpase}.  Their functions range
from providing spatiotemporal organization of cellular materials to tuning biochemical
reactions inside cells \cite{banani2017biomolecular} and improving cellular fitness during
stress \cite{KROSCHWALD2017947,Shineaaf4382}.  To achieve these functions, membraneless
organelles exhibit finely tuned compositional control but the
mechanisms of their operation are incompletely understood.

The interaction of molecules with droplet organelles has recently been described in terms of
a client--scaffold model \cite{banani2016compositional}, where the scaffold is a relatively
stable structure arising from the LLPS process and consists of several protein
and RNA species.  The clients are other molecular species that can be expelled from
or recruited into the droplet comparatively quickly in response to changes in the cellular
environment.  A client may be located transiently in the droplet while the droplet scaffold is effectively static.

Here we use coarse-grained simulations of the client--scaffold model to demonstrate that
superselectivity provides a plausible mechanism for compositional control
in droplet organelles: small variations in the cellular environment can shift the system
across the superselective transition, thus determining whether a species is
found within or outside a droplet organelle.  
Moving from a surface to a 3D binding scaffold, we demonstrate how crowding species can be exploited to manipulate superselective binding.
Membraneless organelles often contain numerous protein types and
RNA species \cite{fong2013whole}, which can act as crowders within the
scaffold.  To the best of our knowledge, this role of molecular crowding in superselective
binding has not been identified before.  We rationalize the behavior using entropic
arguments, systematically evidenced by varying the key parameters.  We also
show that the superselectivity effect is not reliant on details of the
client--receptor interaction model.

\begin{figure}[t]
\includegraphics[width=0.45\textwidth]{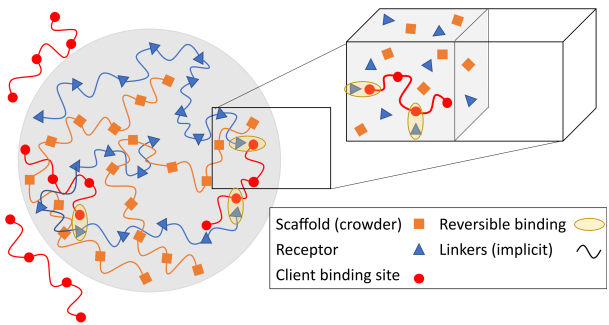}
\caption{Schematic of the simulation setup.  Three types of molecules are simulated, corresponding
to client, receptor and crowder beads.  The beads are connected by flexible implicit linkers.} 
\label{superscheme}
\end{figure}

Our computational model is illustrated schematically in Fig.~\ref{superscheme}.
There are three types of particle, which all occupy vertices on a cubic lattice.
The receptor sites of the scaffold (blue triangles) can reversibly form bonds with
the binding sites on the client molecule (red circles) if they lie on adjacent
sites.  In addition, the scaffold
has crowder beads (orange squares) that occupy a lattice site but have no energetic
interactions with other particles.  A lattice site may host at most one bead,
giving all the beads equal excluded volume. 
In this work, the scaffold is represented by a random distribution of individual receptor and crowder beads for generality.  This approach gives equivalent results to a scaffold of cross-linked polymers for the same receptor and crowder densities, but avoids the results depending on arbitrary details of the scaffold's initialization.  The client is a chain of binding
sites connected by flexible ``implicit'' linkers, as introduced in the Flory
random coil model of Harmon {\em et al.} \cite{harmon2017intrinsically}.  The
linkers have a specified maximum length (in lattice units), limiting the
separation of connected client beads, but the linkers are not
represented explicitly and are treated as occupying negligible volume.

The simulations are performed on a lattice of
$50\times50\times L$ sites with periodic boundary conditions in all directions.
Except where specified, $L = 100$.
Scaffold beads (receptors and crowders) are confined to a slab of thickness 50
sites in the long direction of the cell and periodic in the other two dimensions.
The scaffold-free region
of length $L-50$ lattice sites adjoining the slab represents the cytosol. 
In the client--scaffold model, the client is treated as being far more mobile than the scaffold
structure.  
Thus, the client explores the system while the scaffold remains fixed.  The immobilisation of the scaffold is justified by experimental evidence showing a large variation in retention times of species within membraneless organelles \cite{DITLEV20184666,Kedersha,Boeynaems2019}, with client molecules having significantly higher diffusion rates than scaffold components \cite{dundr2004vivo}. Furthermore, in line with previous work on client-scaffold systems \cite{banani2016compositional,C9SC03191J}, we focus on the dilute regime, where the client concentration is low compared to the scaffold. For simplicity, we work with a single client but, as shown in the Supplemental Material \cite{supplementary}, explicit simulation of multiple clients leads to the same results per client. To ensure statistically significant results, all measurements are averaged over runs with multiple (typically 40) independent scaffold configurations.

The components of membraneless organelles include multivalent species with well defined
binding regions that can accept one ligand each \cite{li2012phase}, as well as
intrinsically disordered proteins or regions of proteins with less specific
attraction \cite{lin2015formation}.  To mimic these, we consider both specific directional
and isotropic interactions between the receptor and client
beads \cite{protter2018intrinsically}.  For isotropic interactions, all client beads
adjacent to receptor beads are bound, while for specific directional interaction,
the receptor bead can form at most one bond.
A directional bond must be broken for either bead to form a new
bond.  We define the bonding interaction energy $-f$ (with $f>0$ for
attractive interactions) with reference to the
temperature, making $f/k T$ the relevant control parameter.
Configuration space is sampled canonically by Monte Carlo steps that alter the client
conformation and bonding arrangements; details of the algorithms are provided in the
Supplementary Material \cite{supplementary}.  For each scaffold snapshot, the typical number of Monte Carlo
sweeps for the client is $10^8$. 

\begin{figure}[t]
\centering
\includegraphics[width=0.48\textwidth]{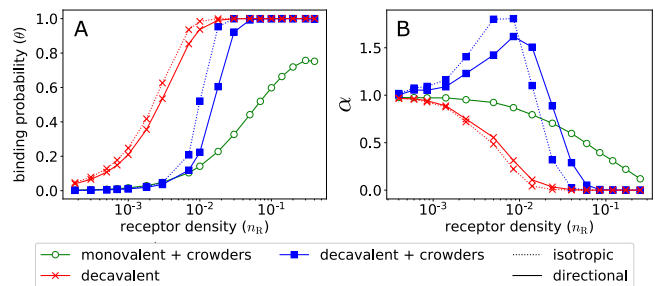}
\caption{Client binding in the isotropic and directional binding models.
The density of crowders (where present) is $0.4$. (A) Binding probability $\theta$
as a function receptor density $n_{{\rm R}}$.  (B) Superselectivity parameter $\alpha$
as a function of receptor density.  The interaction strength between receptors and binding
beads is $f=2kT$ and linkers connecting the binding beads have a length of 5 lattice spacings.}
\label{OOO}
\end{figure}

We define a probability $\theta$ of the client
being bound by at least one bead to the scaffold at a given receptor density $n_{{\rm R}}$ (fraction of lattice sites occupied by receptors). In our dilute 3D system, this probability
is analogous to the fraction of bound particles in studies of binding at two-dimensional
substrates \cite{Dubacheva2015,dubacheva2014superselective,martinez2011designing}.
Figure \ref{OOO}A shows the binding probability as a function of receptor density for
decavalent clients in both the isotropic and directional binding cases and in systems with
and without crowders. The volume ratio between the scaffold and free regions is 1. For reference, we also include corresponding results for a single monovalent client.  The transition from unbound to bound is sharper and occurs within a narrower range of receptor densities for the decavalent client when crowders are present. Furthermore, the sharpness of the increase relative to that of the monovalent client resembles the superselective behavior observed in surface binding \cite{dubacheva2014superselective}. 

To quantify the rate at which binding responds to receptor density, we use the parameter
\begin{equation}
\alpha = \frac{d \: {\rm ln} \: \theta}{d \: {\rm ln} \: n_{{\rm R}}},
\label{eq:alpha}
\end{equation}
as introduced by Martinez-Veracoechea and Frenkel \cite{martinez2011designing} for
nanoparticles binding at a surface.  Here, however, we emphasize that $n_{\rm R}$
is the density (concentration) of receptors in three dimensions.  By construction, $\alpha$
falls monotonically from 1 to 0 with increasing receptor density for any client whose free
energy of binding is independent of that density.  This is the case for a one-bead
monovalent client, as shown by the green circles in Fig.~\ref{OOO}B.  In contrast, if
the binding free energy is sufficiently sensitive to receptor density at low $\theta$,
then $\alpha$ may exhibit a peak above 1, indicating a sharp response
of binding to receptor density and the presence of superselectivity.

For binding at a surface, superselectivity is readily achieved when the binding species is
multivalent.  However, in a 3D scaffold containing only binding sites, this is not always the case. When the volumes of the scaffold and free regions are comparable, we do
not observe superselectivity for any combination of client valency, linker length or binding
strength; the red lines for a decavalent client in Fig.~\ref{OOO}B are typical, with
$\alpha$ never exceeding 1.  Nevertheless, superselectivity can be achieved for both the
isotropic and the directional bonding cases by the introduction of inert crowders, as shown
by the peaks at $\alpha>1$ of the blue lines in Fig.~\ref{OOO}B.  

\begin{figure*}[ht]
\centering
\includegraphics[width=0.95\textwidth]{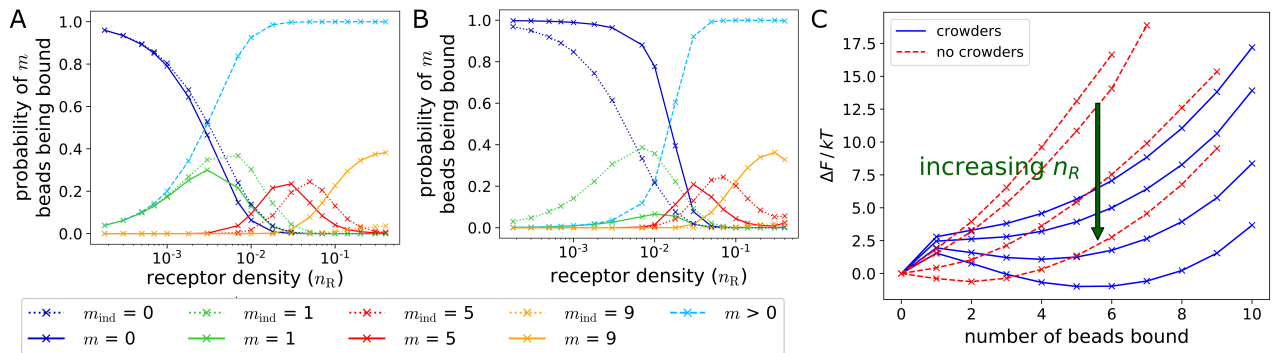}
\caption{Probability of $m$ beads being bound in a decavalent client with linker length 5,
and the probability that $m$ of 10 independent monomers are bound in the
directional binding model. (A) No crowders, (B) crowders at density $0.4$.
The $m>0$ line shows the probability of the chain being bound ($\theta$).  (C)
Change in free energy ($\Delta F \, / \, kT$) for the client to be bound by
$m$ beads in a system with and without crowders, with respect to an unbound client. The interaction energy between receptors
and binding sites is $f=2kT$.}
\label{HIST}
\end{figure*}

As in the surface-binding case, superselectivity in 3D has an entropic origin: binding is
initially suppressed due to an entropic penalty, but the client can then more easily form
many bonds simultaneously \cite{Varilly12a}.  At a surface, the initial decrease in entropy on binding of a flexible
polymeric client arises from both the loss of translational freedom and the restriction
on internal conformations imposed by the surface.  Figure \ref{OOO}B shows that, in a sparse 3D
scaffold, the loss of translational entropy alone may not be sufficient to cause superselectivity, and the conformational entropy must be further controlled by crowding to achieve it.

To quantify this argument, we measure the distribution of the number of bonds that the
decavalent client (with directional binding) forms with the receptors.  The
resulting histograms are shown in Fig.~\ref{HIST}.  For reference, we have also shown the
probability that $m$ of 10 independent monomers are bound in the dilute limit, derived from
simulations of a single monomer and binomial statistics.

The results without crowders are shown in Fig.~\ref{HIST}A.  For clarity, we show
the probabilities only for $m=$ 0, 1, 5 and 9 bound beads.  The probabilities
for $m = 0, 1$ and $5$ are very similar for the multivalent client and the 10 independent
monomers, showing there is little cooperative binding effect.  An increase in the multivalent
probability is only observed for $m = 9$ at a density regime where the chain is already fully
bound and superselectivity is not affected.

The case with crowders is shown in Fig.~\ref{HIST}B.  In contrast to panel (A), we
see an anti-cooperative binding effect for small $m$, relevant at low receptor density
where binding is significantly suppressed for the decavalent client compared to
independent monomers.  This is followed by a similar binding probability for intermediate
$m$, and then a cooperative effect enhancing the binding probability for larger $m$ at
higher receptor density.  This compression of the binding response into a narrower range
of receptor density is the signature of superselectivity.

Figure \ref{HIST}C provides further insight into the thermodynamic origins of the superselectivity.
The change in free energy compared to an unbound client (obtained from the logarithm of the probability distribution) is shown as
a function of the number of bound beads in the decavalent client.  A qualitative difference
between the systems with and without crowders is observed.  In both cases, the
free energy develops a minimum at $m>0$ as the receptor density increases.
However, introducing crowders produces a barrier before the minimum
and the location of the minimum is shifted to larger $m$.  The
barrier suppresses binding of the client to the receptors by a small number of bonds but,
because the entropic penalty only has to be paid once, there is
a rapid increase in the number of bound beads once the barrier is overcome,
as required for superselectivity.

\begin{figure*}[t]
\centering
\includegraphics[width=0.95\textwidth]{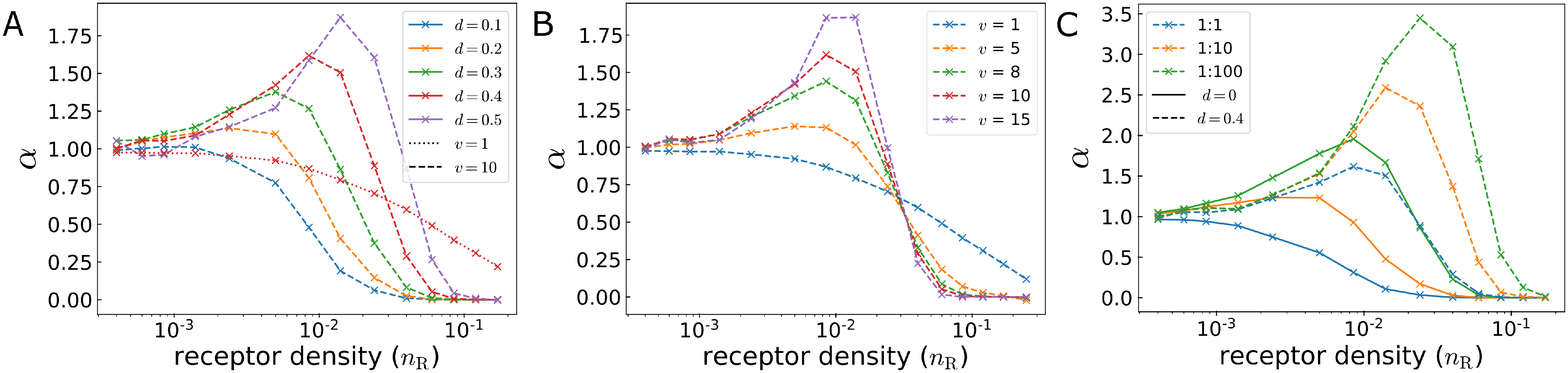}
\caption{Superselective parameter $\alpha$ as a function of receptor density for a range of client parameters in
the directional binding model, varying (A) the density $d$ of crowders
in the scaffold, (B) the valency $v$ of the client, and (C) the ratio of scaffold volume to free space. In (C), the results with and without crowders are compared.  
}
\label{vari}
\end{figure*}

With this understanding in mind, we turn to the question of how the system parameters affect the
degree of superselectivity.  In particular, we expect that superselectivity can be maximized by
control of the entropy loss associated with the early stages of binding, thereby enhancing
cooperative effects beyond that point.  The results of this parameter study are shown in
Fig.~\ref{vari}. Unless it is the parameter being varied, the ratio of scaffold to free volume is 1:1, the client is decavalent with linker length of 5 (lattice units) and the interaction strength is $f/k T = 2$ in a scaffold with crowders at density $d=0.4$.  

First, we vary the crowder density.  Figure \ref{vari}A shows that superselectivity arises
and is progressively enhanced as the density of the crowder beads is increased.
This parameter systematically controls the initial entropic penalty for binding by
restricting the conformations available to the client, allowing superselectivity to be tuned.
The limit of superselective enhancement comes when the crowder density is so high that the client cannot penetrate the scaffold and is primarily bound at the scaffold surface.  Crowders can also occlude receptors by occupying adjacent sites, thereby reducing the effective binding capacity of the scaffold at a given receptor density.

Second, we vary the valency $v$ of the client chain, which in our model is the number of
linked binding sites.  Figure \ref{vari}B shows that
superselectivity increases with valency.  As we have seen, the client's overall binding
free energy must depend sufficiently sensitively on receptor density at the early stages
of binding in order to exhibit superselectivity.  At one extreme in Fig.~\ref{vari}B, the
monomeric client has no internal structure to facilitate such a dependence and
$\alpha$ shows a monotonic decrease.  Increasing
the number of beads in the chain provides the scope for greater entropy loss on entering the
dense 3D scaffold while also increasing the enthalpy of the fully bound state for a given
interaction strength $f$.  Hence, increasing $v$ both raises the initial entropic barrier
and helps to repay the free energy enthalpically, promoting superselectivity.

Third, we study the impact of the free to scaffold volume ratio.  Until now crowding has been necessary to produce any superselective binding.  However, as shown in Fig.~\ref{vari}(C), superselectivity also arises at sufficiently large volume ratio in the absence of crowders.  This is because the entropic penalty required for superselectivity can be introduced by amplifying the loss of translational freedom in the unbound state.  Nonetheless, even in this case, crowders are important because they strongly enhance the degree of superselective binding, thereby sharpening the binding transition.  Microscopy data of various membraneless organelles show that there is a large distribution in the size of droplets and the separation between them, with a typical free to scaffold volume ratio of $O(1000)$, e.g. see \cite{WheelerImage,KedershaImage,KroschwaldImage}.  In the Supplementary Material \cite{supplementary}, we provide a simple statistical mechanical model to extrapolate accurately from computed binding curves to arbitrarily large volumes without further computational expense.

The linker length $l$, interaction strength $f/k T$ and the choice of directional or isotropic bonding affect the extent of superselectivity only weakly.  These dependencies are presented and
rationalized in the Supplementary Material \cite{supplementary} for completeness.  Isotropic binding
introduces a different class of multiply-bound client--receptor configurations compared
to the directional case.  However, the fact that this does not have a strong effect on
superselectivity provides valuable evidence that superselectivity does not rely on details
of the bonding model.

In summary, we have demonstrated the importance of inert crowders in superselective binding of a multivalent chain-like client to a 3D host of binding. The crowders produce an entropic barrier for
the client to enter the 3D scaffold, suppressing binding at low
receptor density.  Once this barrier is overcome, cooperativity derived
from multivalency leads to a sensitive response of binding to increases in the
receptor density. This additional consideration is essential when attempting to generalize from the conventional case of superselectivity at a surface.

The phenomenon of superselectivity may help explain how membraneless organelles exert fine control over the macromolecules that they recruit and expel in the context of simple descriptions like the client--scaffold model \cite{banani2016compositional}. 
The binding receptors in such organelles are held in place by a network of macromolecular backbones, which in themselves constitute part of the background crowder density. Membraneless organelles are known to be susceptible to alteration by changes in the cellular environment, such as pH, salt concentration and glucose availability, or by post translational modifications \cite{DITLEV20184666}. These provide a mechanism by which the scaffold structure and/or the interaction strengths between sites in the membraneless organelle can be tuned, which in turn can lead to expulsion or recruitment of a biomolecular condensate component.

Superselective binding in 3D hosts could be exploited in other supramolecular multivalent structures, such as hydrogels (which have shown potential for tissue engineering \cite{HOFFMAN201218}) and biosensors \cite{hydrogel_biosensor}. Given their polymeric nature, these may be suitable scaffolds to which clients could be attached. Furthermore, supramolecular nanoparticles have been proposed for use in drug delivery \cite{C5NR01526J}, and superselectivity may facilitate different mechanisms for deploying the payload.

Our model for superselectivity in 3D hosts has been minimal in order to capture only the
most essential ingredients of the phenomenon.  Considerable refinement is possible in a
similar spirit, such as more complex client architecture, competing
receptor types \cite{Angioletti-Uberti_competitors} and kinetic control by manipulation
of the free energy profiles.


Raw data relating to this work can be found at \cite{rawdata}.
\begin{acknowledgments}
ATRC is grateful for financial support from the Engineering and Physical Sciences Research Council (UK), Grant EP/L015536/1.
\end{acknowledgments}

\bibliographystyle{apsrev4-1}

\begin{thebibliography}{37}
\bibitem [{raw()}]{rawdata}%
  \BibitemOpen
  \href@noop {} {}\bibinfo {note} {DOI to be added to proofs.}\BibitemShut
  {Stop}%
\end{thebibliography}
%

\end{document}



\title{Supplementary Information: Control of superselectivity by crowding in three-dimensional hosts}



\author{Andrew T. R. Christy}
\affiliation{Department of Chemistry, Durham University, South Road, Durham, DH1 3LE, United Kingdom}
\author{Halim Kusumaatmaja}
\email[]{halim.kusumaatmaja@durham.ac.uk}
\affiliation{Department of Physics, Durham University, South Road, Durham, DH1 3LE, United Kingdom}
\author{Mark A. Miller}
\email[]{m.a.miller@durham.ac.uk}
\affiliation{Department of Chemistry, Durham University, South Road, Durham, DH1 3LE,\ United Kingdom}



\maketitle


\appendix*
\section{Appendix A: Further Simulation Details}

In this work, two types of binding model are employed to describe two common interactions observed within membraneless organelles: isotropic binding (representing generic non-specific attraction) and directional binding (representing binding at mutually specific sites on the two species) \cite{protter2018intrinsically}. In general, the interaction energy between two beads, $i$ and $j$, is given by:
\begin{equation}
\label{hamiltonian}
E_{ij} =  \delta_{ij} \varepsilon_{ij},
\end{equation}
where $\varepsilon_{ij}$ is the interaction strength between beads $i$ and $j$. For a client--receptor pair, $\varepsilon_{ij}$ takes the value $-f$, with $f>0$ corresponding to attractive interactions.  For other combinations of bead types, $\varepsilon_{ij}=0$.  The Kronecker delta $\delta_{ij}$ determines whether or not the two beads are bound. When isotropic bonds are activated, all beads on adjacent lattice sites are automatically bound, and therefore $\delta_{ij}$ is 1 for all adjacent beads and 0 for any pair of beads separated by more than one step on the lattice. In contrast, when directional bonding is applied, each bead can only form one bond. Thus, the calculation of $\delta_{ij}$ requires MC moves for bond breaking and bond formation. The bond formation move involves randomly selecting a client bead neighboring a receptor and attempting a bond formation with an acceptance probability of
\begin{equation}
P^{\rm{acc}} = {\rm{min}}\left[1, \omega e^{-\Delta E/kT}\right] .
\end{equation}
$\Delta E$ is the change in energy of the move. The weighting factor $\omega = N_{\rm u} / (N_{\rm b} + 1)$ is needed to ensure the move obeys detailed balance, where $N_{\rm u}$ and $ N_{\rm b}$ are  the number of unbound and bound client--receptor pairs, respectively. For the bond-breaking move, we can also use a similar method. However, in this case, client--receptor bonds are selected randomly and the weighting factor in the acceptance probability for the selected bond to be broken becomes $\omega = N_{\rm b} / (N_{\rm u} + 1)$  \cite{FrenkelBook}.

\begin{figure}[t]
        \includegraphics[width=0.95\textwidth]{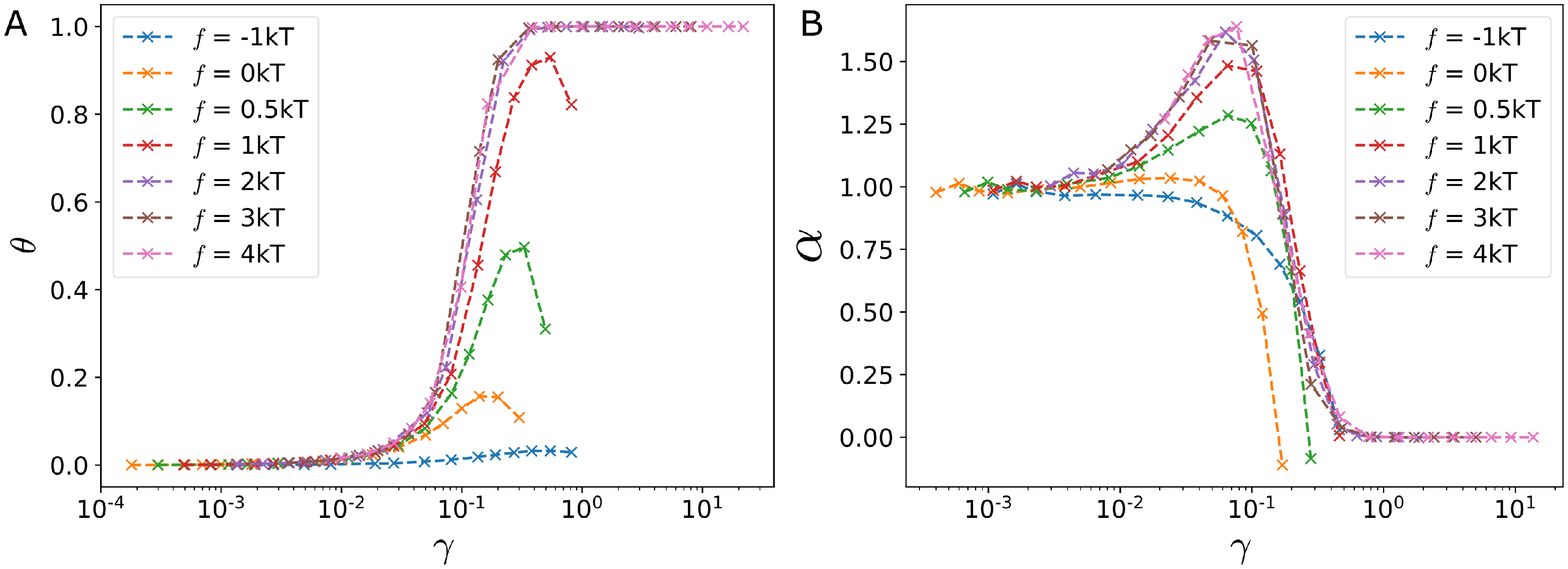}
    \caption{Simulation results for a decavalent client with linker length $l=5$ exploring a system of crowders (at density of 0.4) and receptors in the directional binding case, for an interaction strength between the client beads and receptors ranging from $-1kT$ to $4kT$. (A) Client binding probability $\theta$ as a function of scaled receptor density $\gamma$.  (B) Superselectivity parameter $\alpha=d\ln\theta/d\ln n_{\rm R}$ against scaled receptor density $\gamma=n_{\rm R} e^{-f/kT}$.}
    \label{temp}
    \end{figure}
    
\begin{figure}[t]
        \includegraphics[width=0.95\textwidth]{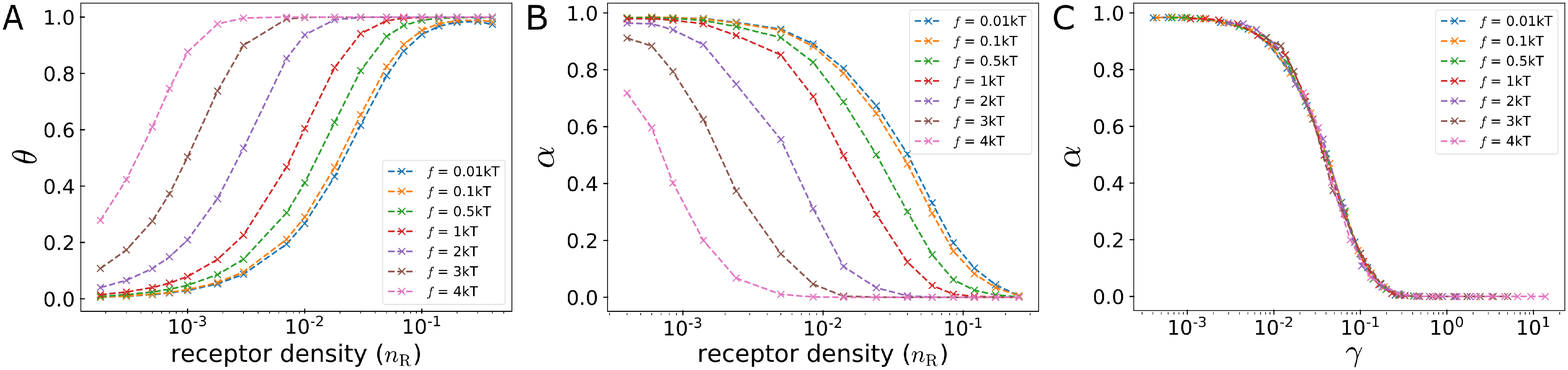}
    \caption{Simulation results for a decavalent client with linker length $l=5$ exploring a system of receptors in the absence of crowders. The strength of the directional interaction between the client and receptor beads range from $0.01kT$ to $4kT$. (A) Client binding probability $\theta$ as a function of receptor density $\gamma$. Superselectivity parameter $\alpha=d\ln\theta/d\ln n_{\rm R}$ against (B) receptor density $n_{\rm R}$ and (C) scaled receptor density $\gamma=n_{\rm R} e^{-f/kT}$.}
    \label{temp_no_crowders}
    \end{figure}    
 
\begin{figure}[t]
        \includegraphics[width=0.4\textwidth]{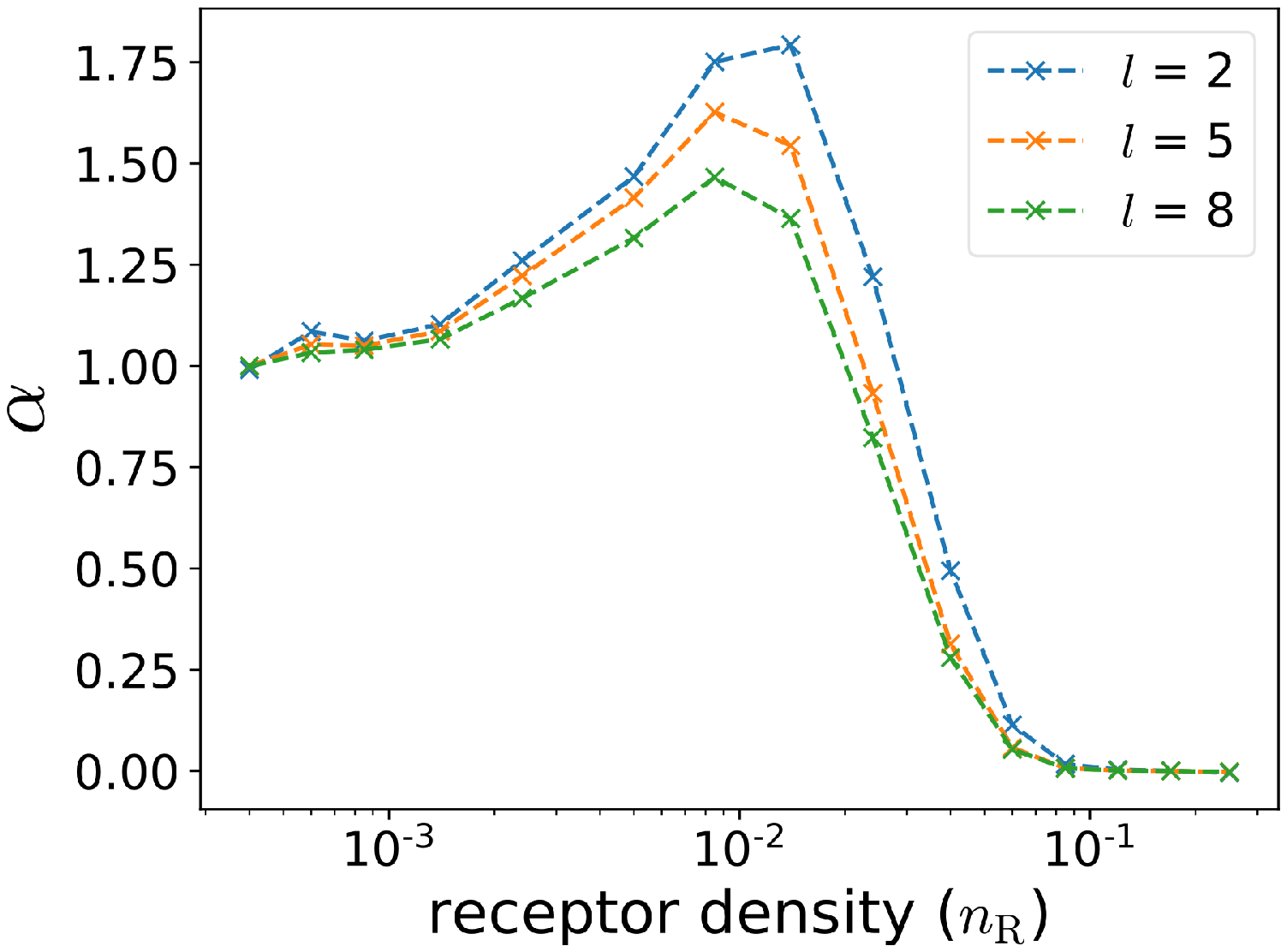}
    \caption{Simulation results for a single decavalent client, with interaction strength of $2kT$, in a system containing crowders at density 0.4, for several linker lengths}
    \label{linker}
    \end{figure}   
 
In the simulations, the client is able to explore the system by reptation, individual bead displacement and teleport moves. In the reptation move, the end bead is deleted and a bead is randomly positioned, within the linker range, at the other end of the chain. Individual bead displacements correspond to moves in which a bead is randomly selected and moved to a new position. Teleport moves involve the deletion of the client and the regrowth from a randomly selected lattice site. This regrowth involves sequentially positioning client beads a random displacement from the previous bead. All these moves are immediately rejected if they result in two beads occupying the same lattice site or the implicit linkers becoming broken. If the move is not limited by either one of these two constraints, it is then accepted with a probability of 
\begin{equation}
P^{\rm{acc}} = {\rm{min}}\left[1,e^{-\Delta E/k T}\right].
\end{equation}
As before, $\Delta E$ is the change in energy of the trial move \cite{FrenkelBook}. We also note that, in the directional binding case, all chain moves that involve a bond being broken are rejected and no bonds are formed in bead displacement moves, in order to obey detailed balance. 

To ensure that the system is fully explored by the client, we typically use $10^8$ MC moves for the directional binding case and $5\times10^8$ for the isotropic binding case.  The ability of the client to explore the system fully was monitored throughout this work, to ensure that the binding occurred all the way through the slab and not predominantly at the interface. Sampling only became an issue when the combined packing fraction of receptors and crowders in the scaffold was 0.8 or greater.  In all cases studied, the binding transition had already occurred at lower densities.  Therefore the inability of the client to penetrate the scaffold at extremely high densities had no impact on the findings in the paper. 
 
 
\section{Appendix B: Varying interaction strength and linker length in the directional binding case}

In this section we provide additional results for the directional binding case. Here we have set the scaffold to free volume ratio to 1:1. Variation in the volume ratio will be further explored in the next section. Figure \ref{temp} shows how superselectivity depends on the interaction strength $f$ for the directional binding case.  Increasing the interaction strength lowers the receptor density at which binding starts.  This enthalpic effect can be accounted for by plotting the probability $\theta$ of binding as a function of the parameter $\gamma = n_{\rm R}e^{-f/kT}$ \cite{martinez2011designing}, where $n_{\rm R}$ is the density of binding beads, defined as the fraction of lattice sites in the scaffold region that the binding beads occupy.  As panel (A) of Fig.~\ref{temp} shows, when plotted in this way, the binding curves overlap for sufficiently strong binding strengths ($f \ge 2kT$).

As the interaction strength weakens below $f=2kT$, the onset of binding is shifted to higher receptor densities.  However, the combination of the crowder beads and the increased number of receptor beads then makes the overall density of the scaffold region very high, thereby increasing the entropic cost for the client to enter the scaffold.  Superselectivity is therefore suppressed by the reduced enthalpic gain and higher entropic penalty for multivalent binding.  This effect is quantified in Fig.~\ref{temp}(B), which shows the superselectivity parameter $\alpha=d\ln\theta/d\ln n_{\rm R}$ as a function of the scaled receptor density $\gamma$.  The peak $\alpha>1$ is unchanged over a wide range of $f$ but is systematically reduced and eventually eliminated when $f$ becomes too small.

To provide additional evidence on the importance of crowders for superselectivity in 3D, we have also carried out simulations for a wide range of interaction strengths in the absence of crowders, from $f = 0.01kT$ to $4kT$. The results are presented in Fig. \ref{temp_no_crowders}. When the volumes of the scaffold and free regions are comparable, we never find superselective binding. Lower interaction strengths do shift the binding transition to higher $n_{\mathrm{R}}$ values, but this shift does not change the steepness of the binding transition. In fact, the binding curves collapse onto a master curve when the receptor density is rescaled through $\gamma$.


We have also studied the impact of varying linker length between the client binding beads. The results are shown in Fig.~\ref{linker}, where it can be seen there is some anti-correlation between superselectivity and the length of the linkers.
Recall that the linkers are represented implicitly as a
maximum tether length between binding beads \cite{harmon2017intrinsically}, so conformations
of the linker sections are not captured explicitly.  The implicit linker
representation also means that the total excluded volume of a client depends only on the
valency (the number of binding beads) and not on the linker length.  However, for short
linkers, the volume-occupying beads are more concentrated in space.  Hence, binding of
even a single bead to a scaffold receptor site places severe constraints on the conformation
that the whole client may adopt.

 \section{Appendix C: Varying the free volume}

The majority of results in this article are for a scaffold system adjacent to an empty cytosol region of equal volume. However, as we have seen in the main text, superselectivity can be observed in the absence of crowders at sufficiently large volume ratio between the free space and the scaffold.
 
\begin{figure}[ht]
        \includegraphics[width=0.9\textwidth]{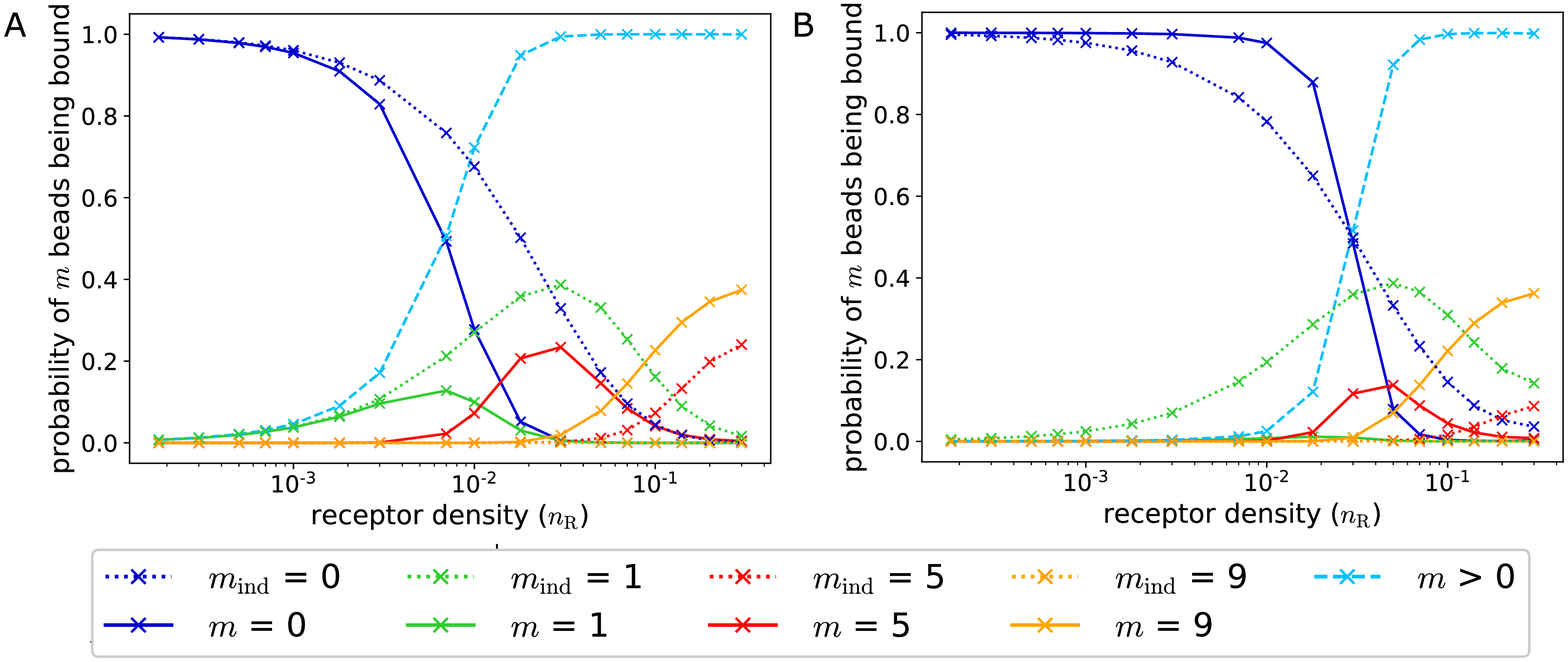}
    \caption{Probability of $m$ beads being bound in a system of parameters $l=5$, $f=2kT$ and with scaffold to free volume ratio of 1:10. The dashed lines are for ten monovalent clients and the solid lines are for a single decavalent client. These plots are for systems (A) without crowders and (B) with crowders at density $d=0.4$.}
    \label{volhist}
    \end{figure}

To probe the entropic origins of this effect, we return to our analysis of multivalent client binding compared to that of monovalent clients. Here, we use a system of scaffold to free volume ratio of 1:10, and study the binding behaviour without and with crowders as a function of the receptor density. The results are shown in Fig.~\ref{volhist}(A) and Fig.~\ref{volhist}(B), respectively. In both cases, a signature of superselective binding is the compression of the binding response into a narrower range of receptor density, due to a suppression at low receptor density and an increase of binding probability at high receptor density. Comparing Fig.~\ref{volhist}(A) and Fig.~\ref{volhist}(B), we can further conclude that this effect is much stronger in the case with crowders.
 
\begin{figure}[ht]
        \includegraphics[width=\textwidth]{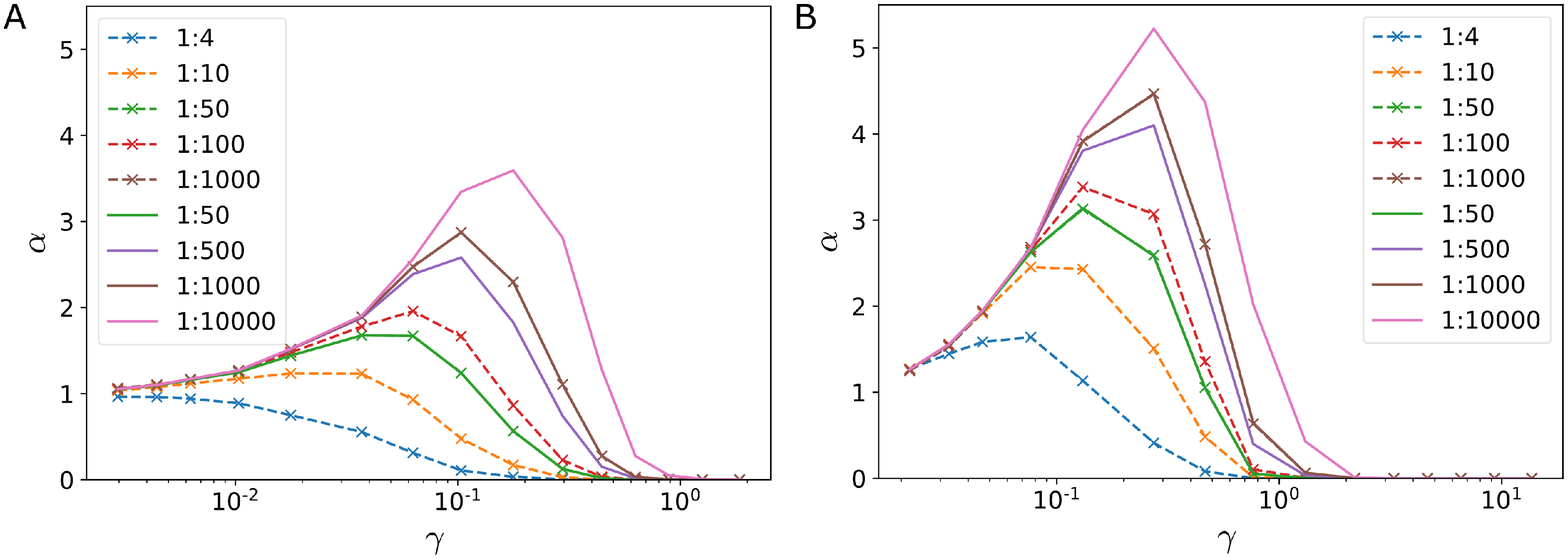}
    \caption{Superselectivity parameter $\alpha$ for a single client in simulation systems with various ratios of scaffold volume to free space, for a scaffold (A) without and (B) with crowders at density $d = 0.4$. Dashed lines are simulation results, and solid lines are predicted curves from Eq.~\ref{extrapolate}. The clients were decavalent with linker length $l=5$ and interaction strength of $f=2kT$.}
    \label{predict}
    \end{figure}

Our direct computer simulations have a practical limitation on the ratio of free space to scaffold volume of $O(1000)$. To demonstrate the dominant effect of crowders in superselective binding for even larger volume ratio, we use a statistical mechanical argument to extrapolate from the binding curve measured in simulations at a computationally accessible volume ratio.  First, we split the partition function of the client molecule into three terms:
\begin{equation}
Q=Q_{\rm b} + Q_{\rm ref} + (V-V_{\rm ref})Q_{\rm u}.
\end{equation}
Here, $Q_{\rm b}$, the partition function for bound configurations in the scaffold and $Q_{\rm ref}$ is the partition function of unbound configurations for a reference volume $V_{\rm ref}$ of free space.  Both these contributions are affected by the presence of the scaffold, since steric interactions between the client and the scaffold will restrict the conformations that the client can adopt, even if the client is not bound to the scaffold.  The only requirement on $V_{\rm ref}$ is that it is large enough to include all unbound conformations of the client that do involve steric interactions with the scaffold.  Finally, $Q_{\rm u}$ is the partition function per unit volume of unbound configurations in the bulk cytosol, far away from any interfaces.  This term captures the internal configurations of the client for a fixed center of mass.  Given this decomposition of $Q$, the probability of binding is
 \begin{equation}
P_{\rm b} = \frac{ Q_{\rm b} } {Q_{\rm b} + Q_{\rm ref} + (V-V_{\rm ref}) Q_{\rm u}}.
\label{Pb}
\end{equation}
The contributions $Q_{\rm b}$ and $Q_{\rm ref}$
do not depend on the actual volume $V$ of the system, while the free-volume term scales linearly with
the volume.  Eq.~(\ref{Pb}) can be rearranged to give
\begin{equation}
\label{extrapolate}
\frac{1}{P_b} = 1 + \frac{Q_{\rm ref}}{Q_{\rm b}} +(V-V_{\rm ref})\frac{Q_{\rm u}}{Q_{\rm b}}.
\end{equation}
Fitting $1/P_{\rm b}$ a function of $V$ from available simulation data at a given receptor density, we can readily infer the values of $(1 + Q_{\rm ref}Q_{\rm b}^{-1}  - V_{\rm ref}Q_{\rm u}Q_{\rm b}^{-1})$ and $Q_{\rm u}Q_{\rm b}^{-1}$ for that receptor density.  This, in turn allows us to extrapolate a point on a binding curve to arbitrarily large volume ratio without carrying out prohibitively expensive simulations.  Repeating the process for each receptor density in a binding curve allows the entire curve to be mapped to a different volume ratio, as shown in Fig.~\ref{predict} for ratios up to 1:10000.  The perfect agreement of the explicit simulations and extrapolated predictions for volume ratios 1:50 and 1:1000 demonstrate the reliability of the calculation.  Importantly, the effect of the crowders remains pronounced and continues to enhance superselectivity even in the limit of large volume ratio between the free space and the scaffold domain.

\section{Appendix D: Simulations with multiple clients}

Throughout this paper, we have presented results using only a single client. Here we demonstrate that these results are representative of the dilute regime in which the clients are at significantly lower concentrations than the scaffold, as has been studied experimentally by Banani et al.~\cite{banani2016compositional} and Jo and Jung \cite{C9SC03191J}, among others. Typically, the client concentrations are several orders of magnitude lower than the scaffold concentrations.

\begin{figure}[t]
        \includegraphics[width=\textwidth]{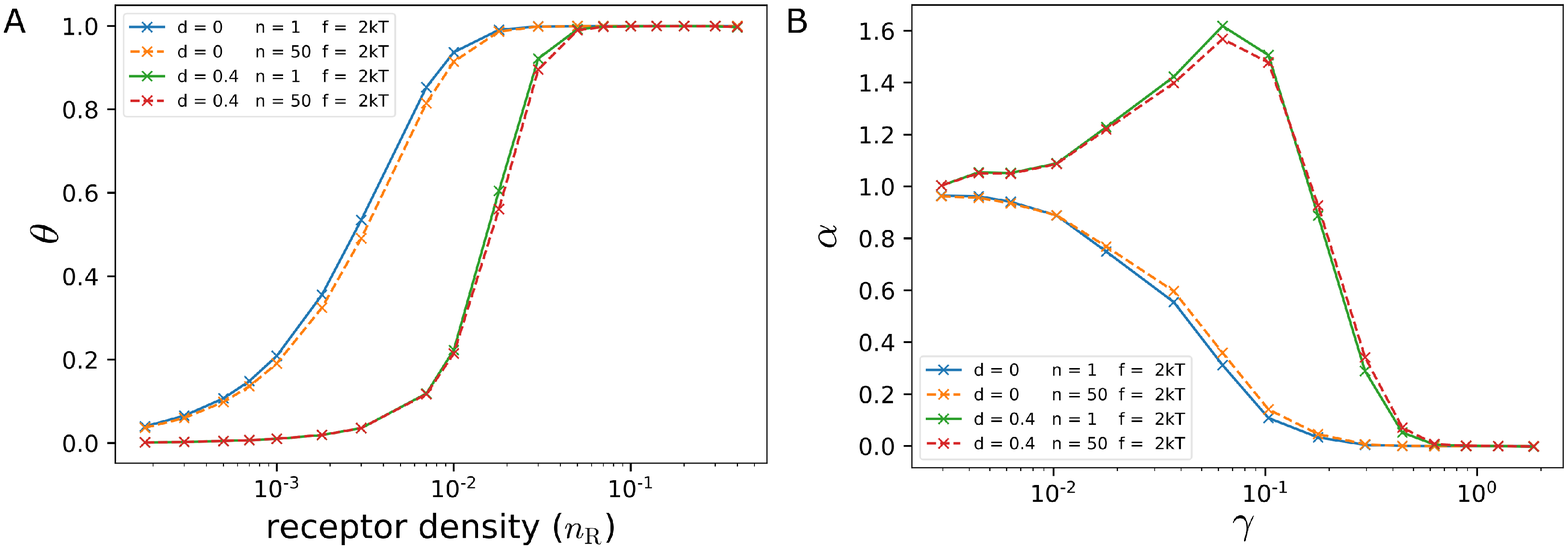}
    \caption{Results for simulations of 50 clients within the system, where (A) shows the probability of binding at various receptor densities and (B) shows the variation in $\alpha$ with $\gamma$. }
    \label{multi}
    \end{figure}

Fig.~\ref{multi} compares the simulation results using 1 and 50 decavalent clients for the directional binding case, corresponding to a client volume fraction of $\phi=8 \times 10^{-5}$ and $\phi = 4 \times 10^{-3}$ respectively. The fraction of bound clients and the superselectivity behaviour observed are essentially identical. We only start seeing significant deviation for 100 clients, corresponding to a client fraction of $\phi = 8 \times 10^{-3}$.

The structure and composition of membranelles organelles remain a very active area of research, and the proportions of client and scaffold molecules that form stable droplets depend strongly on the types of membranelles organelles \cite{promiscuous,Rosen2020}. We expect higher client concentrations to alter the scaffold structure \cite{nakashima2019biomolecular}, and this is an interesting and important area for future work.

 \begin{figure*}[t]
        \includegraphics[width=0.93\textwidth]{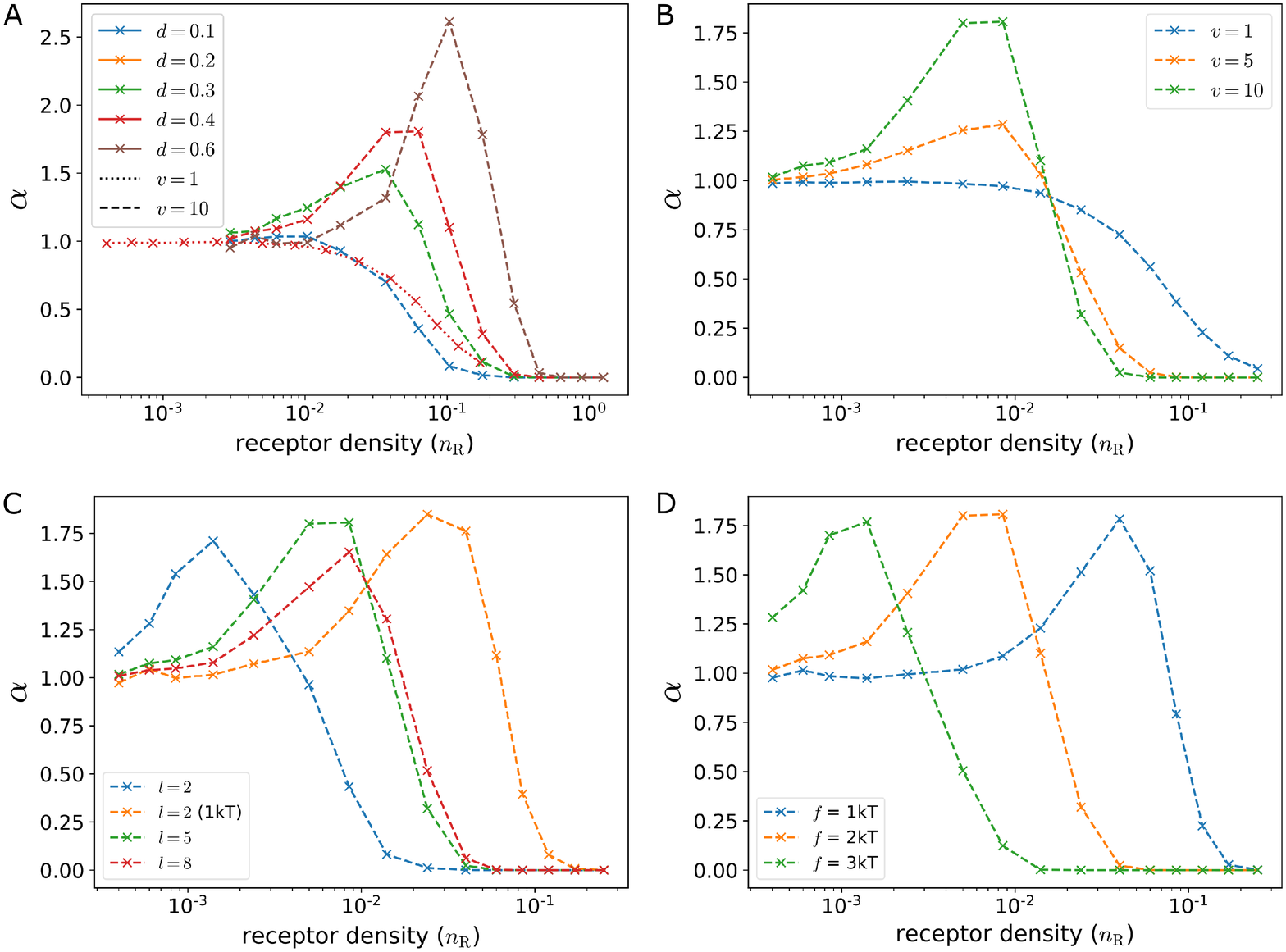}
    \caption{The superselectivity parameter $\alpha$ as a function of $n_{{\rm R}}$ for a variety of parameters: (A) the density $d$ of crowders in the receptor slab, (B) the valency of the client $v$, (C) the linker length of the client $l$, and (D) the interaction strength between the client and receptors $f$. In each of these plots the variables are $v=10$, $l=5$, $f=2kT$ and $d=0.4$, unless stated otherwise. All results in this figure are for the case of isotropic binding. }
        \label{variables}
\end{figure*}
 
\section{Appendix E: Isotropic binding}
 
\begin{figure}[t]
        \includegraphics[width=\textwidth]{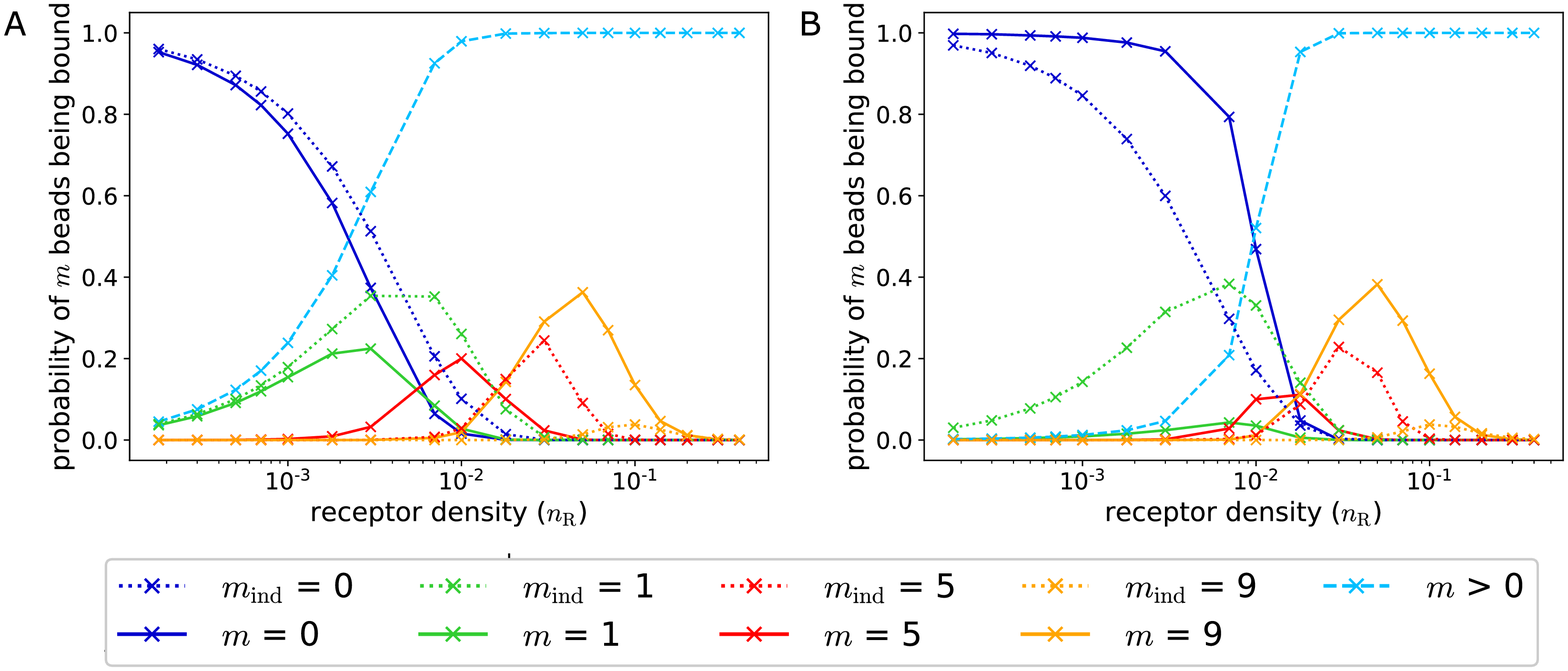}
    \caption{Probability of $m$ beads being bound in a system of parameters $l=5$ and $f=2kT$. The dashed lines are for ten monovalent clients and the solid lines are for a single decavalent client. These plots are for systems (A) without crowders, (B) with crowders at density $d=0.4$.}
    \label{isohist}
    \end{figure} 

The results for the isotropic binding case align well with those of the directional binding case, as can be seen in Fig.~\ref{variables}. We observe that superselectivity increases with increasing crowder density and client valency in panels (A) and (B), respectively. However, deviations from the directional binding case can be observed for variations in linker length and binding strength, as illustrated in panels (C) and (D). The key new effect for isotropic binding is due to the possibility of forming bonds between multiple beads of the client and a single receptor bead, which is not allowed in the directional model. 

By forming multiple bonds, the effective density of receptor sites is increased and the client can more easily overcome the entropic cost of entering the 3D host. An extreme case is shown in Fig.~\ref{variables}(C) for $l=2$, where the receptor beads are surrounded by a client, resulting in the peak in $\alpha$ at very low $n_{{\rm R}}$. Such behavior requires the client to adopt a compact structure. Correspondingly, as further shown in Fig.~\ref{variables}(C), we can delay the peak in $\alpha$ by increasing the entropic barrier by using longer linker length, or by decreasing the enthalpic gain by employing weaker interaction strength (comparing $f=2\,kT$ and $f=1\,kT$ for $l =2$).

Similar behavior is also observed in Fig.~\ref{variables}(D) for $l = 5$ when we vary $f$ from $1\,kT$ to $3\,kT$. These shifts in $n_{{\rm R}}$ values for the peaks of $\alpha$ differentiate the isotropic from the directional binding case. For the latter, the shifts are not observed upon variation in linker length. The shifts for variations in interaction strength also cannot be captured by the parameter $\gamma$ introduced in the previous section. In fact, since the client can form multiple bonds, the use of $\gamma$ instead of $\theta$ for the binding probability plots is generally not helpful.

Figure \ref{isohist} shows the probability of $m$ beads being bound for the isotropic binding case, in systems both with and without crowders. As with the directional binding case, the probability of one bead being bound is greatly suppressed for a multivalent client compared to monomers in a system with crowders, but not when crowders are absent. The main difference between the directional and isotropic binding cases is that the cooperative effect sets in at lower $n_{{\rm R}}$. This can be seen by comparing the lines corresponding to 5 beads being bound, for the decavalent and monovalent clients. This is likely a result of the client being able to form many more bonds in the isotropic binding case, compared to directional binding. Isotropic binding also reduces the competition for binding to a given receptor as multiple client beads can bind to the same receptor, unlike in the directional binding case. 

Overall, the earlier onset of cooperativity strengthens superselective effect where crowders are present. This can be observed by comparing the magnitude of the peaks in Fig.~\ref{variables} and Fig.~4 of the main text. However, this change in cooperativity is not sufficient to introduce superselective binding in the case where crowders are absent. We also emphasize that the possibility for a client to form multiple bonds with a receptor bead differentiate the isotropic binding case from the directional binding scenario. These two cases cannot be mapped onto each other through a simple rescaling of the interaction energies.  Nevertheless, the trends in superselective behavior with respect to crowder density are analogous for the two binding models, demonstrating that the introduction or enhancement of superselectivity by crowders does not depend on the details of the bonding.

\bibliographystyle{apsrev4-1}
\bibliography{paperref}